\documentclass[preprintnumbers]{revtex4}
\usepackage{amssymb}
\usepackage{amsmath}
\usepackage{graphicx}
\usepackage{calligra}
\usepackage{mathrsfs}
\usepackage{dcolumn}
\usepackage{bm}
\usepackage{subfigure}
\usepackage{color}
\usepackage{CJKutf8}
\numberwithin{equation}{section}
\numberwithin{figure}{section}

\begin{document}
\title{Thermodynamic relation on rotating charged black strings with arbitrary cosmological constant}
\author{Hai-Long Zhen$^{1,2}$, Meng-Sen Ma$^{1,2}$, Huai-Fan Li$^{2,3}$, Jian-Hua Shi$^{1,2}$, and Yu-Bo Ma$^{1,2}$}
\thanks{\emph{e-mail: yuboma.phy@gmail.com}(corresponding author)}

\address{$^1$Department of Physics, Shanxi Datong University, Datong 037009, China\\
$^2$Institute of Theoretical Physics, Shanxi Datong University, Datong, 037009, China\\
$^3$College of General Education, Shanxi College of Technology, Shuozhou 036000, China\\}

\begin{abstract}

The Goon-Penco (GP) relation was  investigated on rotating charged black strings with an arbitrary cosmological constant. It has been demonstrated that the GP relation retains its form in the context of spacetimes described by cylindrical coordinates. In addition, the GP relation is derived in scenarios where the energy state parameters (including angular momentum $J$ and charge $Q$, etc.) are expressed as functions of the perturbation parameter $\eta$. This finding indicates that the GP relation is not only valid for spacetimes described by spherically symmetric coordinates, but also prevalent for non-spherically symmetric spacetimes, such as cylindrical coordinates. Therefore, the present study demonstrated that the GP relation is universal for spacetimes with arbitrary cosmological constants, irrespective of the adopted coordinate system.

\end{abstract}

\maketitle

\section{Introduction}\label{one}

Black holes represent a distinctive physical system, manifesting characteristics of both strong gravitational fields and thermodynamic systems. As early as the 1970s, prominent physicists such as Hawking and Bekenstein established the four laws of thermodynamics of black holes \cite{HSW,BJD}, thereby establishing the theoretical foundation for this field of study. The in-depth study of the thermodynamic properties of black holes is of dual significance. Firstly, it has been demonstrated to facilitate a more profound comprehension of the fundamental properties of black holes. Secondly, it provides novel perspectives for exploring the internal microscopic states of black holes. In the context of contemporary black hole thermodynamic research, such as the intrinsic correlation between the state parameters of black holes, the specific thermodynamic effects, as well as the similarities and differences between black hole thermodynamics and ordinary thermodynamic systems must be examined. These studies will contribute to our comprehensive understanding of black hole physics.

In the context of quantum gravity theory, the Weak Gravity Conjecture (WGC) was proposed as a criterion \cite{C0509212}. The present conjecture posits that, within any system comprising a particle with mass $M$ and charge $Q$, gravity must be the weakest force. This can be expressed as $\frac{Q}{M}>1$. moreover, it has been demonstrated that extremal black holes exhibit $\frac{Q}{M}=1$ \cite{N0601001}. In order to reconcile the WGC with extremal black hole solutions, higher-derivative operators with specific coefficients are introduced, leading to the formulation of the black hole Weak Gravity Conjecture \cite{Y0901}. It is noteworthy that the generalized WGC, incorporating higher-derivative corrections, remains valid even for evaporating black holes emitting uncharged particles \cite{T084019,D035003}. The WGC plays a crucial role in theoretical physics by providing correction terms that naturally eliminate naked singularities \cite{C08546}. This significant feature establishes a profound connection: the absence of naked singularities inherently validates the WGC inequality \cite{C08546}. Goon and Penco investigated the universal relation between corrections to entropy and extremality of black holes under the perturbation to advance our understanding of the WGC \cite{G101103}. The GP relation demonstrates that the extremality relations of black holes exhibit WGC-like behavior \cite{G101103},
\begin{align}\label{1.1}
\frac{\partial {{M}_{ext}}(Q,\eta )}{\partial \eta }&=-\underset{M\to {{M}_{ext}}}{\mathop{\lim }}\,T{{\left( \frac{\partial S(M,Q,\eta )}{\partial \eta } \right)}_{Q,M}}.
\end{align}
The Eq. (\ref{1.1}) is a universe relation connecting the derivatives of mass-like function with the entropy function with respect to the perturbation  $\eta$. Based on the relation proposed by Goon and Penco, considerable progress has been made \cite{J025305,S11161,P08530,M13784,J139149,J17014,C116686,L08527,S115279,A07079,H11535,M09654,L00015,BL081901,J169168,Y15551,S250412,Y035106}.

The extant studies on the GP relation are principally constrained to the spacetime delineated by the spherical coordinate. The angular momentum $J$ of a rotating black hole is widely regarded as a fixed state parameter, independent of the perturbation parameter $\eta$ \cite{J17014}. However, there is a paucity of systematic studies examining the applicability of this relation when considering two cases: (1) black holes with rotational parameters whose angular momentum $J$ behaves as a function of $\eta$; and (2) thermodynamic quantities of horizon interfaces described by non-spherical coordinate systems (e.g., cylindrical, etc.) have a dependence on $\eta$. These two scenarios introduce novel theoretical challenges that necessitate exploration of the Goon-Penco relation between thermodynamic quantities. In order to verify the generalizability of the Goon-Penco relationship, there is an urgent need for in-depth studies in the following areas:

The paper is arranged as follows: In Sec. \ref{two}, a universal relation between the energy and entropy of high-dimensional, spherically symmetric spacetime under perturbation are reviewed. The universal Goon-Penco relation corresponding to thermodynamic quantities at the black hole and cosmological horizon is derived. In Sec. \ref{three}, the Goon-Penco relation for the non-spherically symmetric Rotating Charged Black String is derived under the condition that the thermodynamic quantities at the spacetime interface satisfy the first law of thermodynamics. The GP relation under different constraints is obtained when the thermodynamic quantities of the Black String are all functions of the perturbation parameter $\eta$. In conclusion, a brief summary is presented in Sec. \ref{four}

\section{Higher-dimensional RN(A)dS black holes: general case} \label{two}

To extend to our anaysis to the non-spherically symmetric case, we review our calculation to a more general case, this is, higher-dimensional, spherically spacetime. The action of higer-dimensional Ads black holes is expressed as
\begin{align}\label{2.1}
I&=\frac{1}{16\pi }\int{{{d}^{n+2}}}x\sqrt{g}(R-{{F}_{\mu \nu }}{{F}^{\mu \nu }}-2(1+\eta )\Lambda ),
\end{align}
Thus,the metric function is expressed as \cite{M13784,R0112253,R07225}
\begin{align}\label{2.2}
\Delta (r)&=1-\frac{{{\omega }_{n}}M}{{{r}^{n-1}}}+\frac{n\omega _{n}^{2}{{Q}^{2}}}{8(n-1){{r}^{2n-2}}}-\frac{1+\eta }{{{l}^{2}}}{{r}^{2}},~~{{\omega }_{n}}=\frac{16\pi }{nVol({{S}^{n}})}.
\end{align}
Where $Q$ is the electric$/$magnetic charge of Maxwell field. For general $M$ and $Q$, the equation $\Delta (r)=0$ may have four real roots. Three of them are real: the largest one is the cosmological horizon $r_c$, the smallest is the inner(Cauchy) horizon of black hole, the middle one is the outer horizon $r_+$ of the black hole. And the fourth is negative and has no physical meaning. The classification of the RNdS solution has been made in Ref \cite{L9203018}. Some thermodynamic quantities associated with the cosmological horizon and black hole horizon are
\begin{align}\label{2.3}
M&=\pm \frac{r_{+,c}^{n-1}}{{{\omega }_{n}}}\left( 1-\frac{(1+\eta )r_{+,c}^{2}}{{{l}^{2}}}+\frac{n\omega _{n}^{2}{{Q}^{2}}}{8(n-1)r_{+,c}^{2n-2}} \right),~~{{\phi }_{+,c}}=\pm \frac{n{{\omega }_{n}}Q}{4(n-1)r_{+,c}^{n-1}}, \notag \\
{{T}_{+,c}}&=\pm \frac{1}{4\pi {{r}_{+,c}}}\left( (n-1)-(n+1)\frac{(1+\eta )r_{+,c}^{2}}{{{l}^{2}}}-\frac{n\omega _{n}^{2}{{Q}^{2}}}{8r_{+,c}^{2n-2}} \right),~~{{S}_{+,c}}=\frac{Vol({{S}^{n}})}{4}r_{+,c}^{n}.
\end{align}
Through Eq. (\ref{2.3}), we obtain
\begin{align}\label{2.4}
{{\left( \frac{\partial M}{\partial {{S}_{+,c}}} \right)}_{l,Q}}&=\frac{\partial M}{\partial {{r}_{+,c}}}\frac{\partial {{r}_{+,c}}}{\partial {{S}_{+,c}}}=\pm \left( (n-1)\frac{r_{+,c}^{n-2}}{{{\omega }_{n}}}-\frac{(1+\eta )(n+1)r_{+,c}^{n}}{{{\omega }_{n}}{{l}^{2}}}-\frac{n{{\omega }_{n}}{{Q}^{2}}}{8(n-1)r_{+,c}^{n-2}} \right)\frac{4}{Vol({{S}^{n}})nr_{+}^{n-1}} \notag \\
&=\pm \frac{1}{4\pi {{r}_{+,c}}}\left( (n-1)-(n+1)\frac{(1+\eta )r_{+,c}^{2}}{{{l}^{2}}}-\frac{n\omega _{n}^{2}{{Q}^{2}}}{8r_{+,c}^{2n-2}} \right)={{T}_{+,c}},
\end{align}

When the charge $Q$ is held fixed, while the energy, $M=M_0$, of the spacetime is treated as a invariant quantity(the parameter $M_0$ must satisfy the condition for the coexistence of the black hole and cosmological horizon). From Eq. (\ref{2.3}), we obtain
\begin{align}\label{2.5}
{{\left( \frac{\partial {{M}_{0}}}{\partial \eta } \right)}_{l,Q,{{S}_{+,c}}}}&=\mp \frac{r_{+,c}^{n+1}}{{{\omega }_{n}}{{l}^{2}}},
\end{align}
when the charge $Q$ and the energy, $M=M_0$, of the spacetime are held fixed, from Eq. (\ref{2.3}), we obtain
\begin{align}\label{2.6}
0&=\pm \left( (n-1)\frac{r_{+,c}^{n-2}}{{{\omega }_{n}}}-\frac{(1+\eta )(n+1)r_{+,c}^{n}}{{{\omega }_{n}}{{l}^{2}}}-\frac{n{{\omega }_{n}}{{Q}^{2}}}{8(n-1)r_{+,c}^{n-2}} \right)d{{r}_{+,c}}\mp \frac{r_{+,c}^{n+1}}{{{\omega }_{n}}{{l}^{2}}}d\eta,
\end{align}
through Eq. (\ref{2.6}), we obtain
\begin{align}\label{2.7}
{{\left( \frac{\partial {{S}_{+,c}}}{\partial \eta } \right)}_{l,Q,M}}&=\pm \frac{r_{+,c}^{(n+1)}}{{{\omega }_{n}}{{l}^{2}}}\frac{1}{{{T}_{+,c}}}.
\end{align}
By comparing Eqs. (\ref{2.5}) with (\ref{2.7}), The relation between energy and entropy proposed by Goon and Penco can be derived as follows
\begin{align}\label{2.8}
{{\left( \frac{\partial {{M}_{0}}}{\partial \eta } \right)}_{l,{{S}_{+,c}},Q}}&=\underset{M\to {{M}_{0}}}{\mathop{\lim }}\,\mp {{T}_{+,c}}{{\left( \frac{\partial {{S}_{+,c}}}{\partial \eta } \right)}_{l,Q,M}}.
\end{align}

It has been demonstrated that the thermodynamic quantities associated with both horizons of the spacetime satisfy the first law of thermodynamics.
\begin{align}\label{2.9}
dM&={{T}_{+,c}}d{{S}_{+,c}}+{{\phi }_{+,c}}dQ,
\end{align}
with
\begin{align}\label{2.10}
{{T}_{+,c}}&=\pm {{\left( \frac{\partial M}{\partial {{S}_{+,c}}} \right)}_{l,Q}},~~{{\phi }_{+,c}}=\pm {{\left( \frac{\partial M}{\partial Q} \right)}_{l,{{S}_{+,c}}}}.
\end{align}
According to Eq. (\ref{2.3}), the position $r_{+,c}$ of the black hole and cosmological horizon are functions of the energy $M$, charge $Q$, and perturbation parameter $\eta$. Given the charge $Q$, the variation of the energy $M$ induced by the perturbation parameter $\eta$ is described by the following equation:
\begin{align}\label{2.11}
\frac{dM}{d\eta }&={{\left( \frac{\partial M}{\partial \eta } \right)}_{l,{{S}_{+,c}},Q}}+{{\left( \frac{\partial M}{\partial {{S}_{+,c}}} \right)}_{l,Q}}{{\left( \frac{\partial {{S}_{+,c}}}{\partial \eta } \right)}_{l,Q}}={{\left( \frac{\partial M}{\partial \eta } \right)}_{l,{{S}_{+,c}},Q}}\pm {{T}_{+,c}}{{\left( \frac{\partial {{S}_{+,c}}}{\partial \eta } \right)}_{l,Q}},
\end{align}

When the energy $M=M_0$ of the black hole are held fixed, Eq. (\ref{2.11}) obey the following equation:
\begin{align}\label{2.12}
{{\left( \frac{\partial {{M}_{0}}}{\partial \eta } \right)}_{l,{{S}_{+,c}},Q}}&=\underset{M\to {{M}_{0}}}{\mathop{\lim }}\,\mp {{T}_{+,c}}{{\left( \frac{\partial {{S}_{+,c}}}{\partial \eta } \right)}_{l,Q,M}}.
\end{align}

By utilizing the thermodynamic first law satisfied at both horizons and incorporating the relationship between spacetime energy $M$ and perturbation parameter $\eta$, the energy-entropy relation proposed by Goon and Penco is derived. This approach offers two significant advantages: first, it circumvents complex mathematical computations, and second, it establishes a novel, simplified pathway for investigating the Goon-Penco relation in complex spacetimes. In the context of non-spherically symmetric spacetimes, where the state function of energy $M$ incorporates supplementary parameters (e.g., angular momentum $J$), which in turn are functions of $\eta$. Consequently, advancements in the examination of the Goon-Penco relation have been constrained. This is primarily due to the fact that the energy state function's explicit dependence on the parameter $\eta$ introduces additional complications. In the subsequent section, we employ our novel methodology to methodically examine the Goon-Penco relation in such non-spherically symmetric spacetimes.

\section{Rotating Charged Black Strings} \label{three}

For a four-dimensional spacetime with cylindrical or toroidal horizons, the line element can be expressed as \cite{J9404041,J9511188,H09375}:
\begin{align}\label{3.1}
d{{s}^{2}}&=-f(r){{(\Xi dt-ad\varphi )}^{2}}+{{r}^{2}}{{\left( \frac{a}{{{l}^{2}}}dt-\Xi d\varphi  \right)}^{2}}+\frac{d{{r}^{2}}}{f(r)}+\frac{{{r}^{2}}}{{{l}^{2}}}d{{z}^{2}}.
\end{align}

The metric function $f(r)$ is determined by
\begin{align}\label{3.2}
f(r)&=\frac{{{r}^{2}}}{{{l}^{2}}}-\frac{m}{r}+\frac{{{q}^{2}}}{{{r}^{2}}},~~{{\Xi }^{2}}=1+\frac{a}{{{l}^{2}}},
\end{align}
and the gauge potential admits the the following form
\begin{align}\label{3.3}
A&=-\frac{q}{r}(\Xi dt-ad\varphi ).
\end{align}
The relevant thermodynamic potentials are determined as
\begin{align}\label{3.4}
M&=\frac{1}{16\pi l}(3{{\Xi }^{2}}-1)m,~~T=\frac{3r_{+}^{4}-{{q}^{2}}{{l}^{2}}}{4\pi r_{+}^{3}{{l}^{2}}\Xi },~~S=\frac{r_{+}^{2}\Xi }{4l},~~\Omega =\frac{a}{\Xi {{l}^{2}}},\notag \\
J&=\frac{3}{16\pi l}\Xi am,~~U=\frac{q}{{{r}_{+}}\Xi },~~Q=\frac{q\Xi }{4\pi l},~~m=\frac{r_{+}^{3}}{{{l}^{2}}}+\frac{{{q}^{2}}}{{{r}_{+}}}.
\end{align}
The event horizon $r_{+}$ corresponds to the largest root of the metric function in Eq.(\ref{3.2}), satisfying $f(r_{+})=0$. This condition leads to the mass parameter $m$, given in Eq. (\ref{3.4}), in terms of the horizon radius $r_+$, the charge paramete $q$ and the AdS radius $l$. When expressing the mass per unit volume of black string horizon $M$ in terms of the entropy $S$, charge $Q$, and angular momentum $J$ per unit volume of black string horizon (i.e., $M(S,Q,J)$), the first law of black hole thermodynamics takes the following form:
\begin{align}\label{3.5}
dM&=TdS+\Omega dJ+UdQ.
\end{align}
In the presence of a cosmological constant, $\Lambda$, a natural candidate for thermodynamic pressure emerges as follows:
\begin{align}\label{3.6}
{{P}_{0}}&=-\frac{\Lambda }{8\pi }=\frac{3}{8\pi {{l}^{2}}}.
\end{align}

It is proposed that the thermodynamic volume of the black string per unit horizon volume, $V$, be defined as the thermodynamic variable conjugate to ${P}_{0}$. By allowing the pressure ${P}_{0}$, the first law of black hole thermodynamics should then be modified to \cite{A07079}
\begin{align}\label{3.7}
dM=TdS+\Omega dJ+UdQ+Vd{{P}_{0}}+{{\left( \frac{\partial M}{\partial \eta } \right)}_{S,J,Q,{{P}_{0}}}}d\eta.
\end{align}
According to ref \cite{H09375}, when the spacetime perturbation parameter, $\eta$=0, the expression for the mass, denoted by $M$, is given in terms of the thermodynamic variables, denoted by $S, Q, J$, and $P$,
\begin{align}\label{3.8}
M&=\frac{1}{{{2}^{3/4}}\sqrt{3S}\sqrt[4]{P\pi }{{[32{{(\pi P)}^{3/2}}\sqrt{2}{{J}^{2}}S+\psi )}^{3/2}}}[32{{(\pi P)}^{3/2}}\sqrt{2}{{J}^{2}}S\psi +36854{{P}^{4}}{{S}^{8}}\notag \\
&+55296\pi {{P}^{3}}{{Q}^{2}}{{S}^{6}}+32{{\pi }^{3}}P{{S}^{2}}(64{{J}^{4}}{{P}^{2}}+243{{Q}^{6}})+31104{{\pi }^{2}}{{P}^{2}}{{Q}^{4}}{{S}^{4}}+729{{\pi }^{4}}{{Q}^{8}}],
\end{align}
where
\begin{align}\label{3.9}
{{\psi }^{2}}&=110592{{P}^{4}}{{S}^{8}}+165888\pi {{P}^{3}}{{Q}^{2}}{{S}^{6}}+32{{\pi }^{3}}PS(64{{J}^{4}}{{P}^{2}}+729{{Q}^{6}})+93312{{\pi }^{2}}{{P}^{2}}{{Q}^{4}}{{S}^{4}}+2197{{\pi }^{4}}{{Q}^{8}}.
\end{align}
According to Eq.(\ref{3.7}), differentiating with respect to $P$ results in the thermodynamic volume, expressed in terms of the thermodynamic variables
\begin{align}\label{3.10}
V&={{\left( \frac{\partial M}{\partial P} \right)}_{S,J,Q}}.
\end{align}
In terms of thermodynamic variables($S,Q,J,P$), the temperature is
\begin{align}\label{3.11}
T&={{\left( \frac{\partial M}{\partial S} \right)}_{P,J,Q}}.
\end{align}
The angular velocity is
\begin{align}\label{3.12}
\Omega &={{\left( \frac{\partial M}{\partial J} \right)}_{S,P,Q}},
\end{align}
and the electric potential is
\begin{align}\label{3.13}
U&={{\left( \frac{\partial M}{\partial Q} \right)}_{S,P,J}}.
\end{align}

When perturbation parameter $\eta{\neq}0$ in spacetime, $\Lambda{\Rightarrow}{\Lambda}(1+\eta)$ \cite{G101103,M13784,J17014,A07079}, Assuming that $\eta$ is a small parameter and ${\eta}{\rightarrow}0$, the action reduces to its uncorrected form.
\begin{align}\label{3.14}
P&=\frac{3(1+\eta )}{8\pi {{l}^{2}}}={{P}_{0}}(1+\eta ).
\end{align}
According to Eq. (\ref{3.4}), when treating $\eta$ as a constant, the thermodynamic relationships of Eqs. (\ref{3.5})-(\ref{3.13}) are expressed in the same form, except that the coefficients in the aforementioned equations are replaced as follows:
\begin{align}\label{3.15}
P&=\frac{3(1+\eta )}{8\pi {{l}^{2}}}={{P}_{0}}(1+\eta ),~~J=J(\eta ),~~Q=Q(\eta ),~~S=S(\eta ).
\end{align}
Therefore, the state parameter of the energy, denoted by $M$, in Eq. (\ref{3.8}) is expressed as $M(S, P_0, J(\eta), Q(\eta), \eta)$. When $\eta$ is subject to perturbation, from Eqs. (\ref{3.7}) and (\ref{3.8}), we obtain
\begin{align}\label{3.16}
\frac{dM}{d\eta }&=\left( \frac{\partial M}{\partial \eta } \right)+\left( \frac{\partial M}{\partial S} \right)\left( \frac{\partial S}{\partial \eta } \right)+\left( \frac{\partial M}{\partial Q} \right)\left( \frac{\partial Q}{\partial \eta } \right)+\left( \frac{\partial M}{\partial J} \right)\left( \frac{\partial J}{\partial \eta } \right)+\left( \frac{\partial M}{\partial {{P}_{0}}} \right)\left( \frac{\partial {{P}_{0}}}{\partial \eta } \right) \notag \\
&=\left( \frac{\partial M}{\partial \eta } \right)+T\left( \frac{\partial S}{\partial \eta } \right)+U\left( \frac{\partial Q}{\partial \eta } \right)+\Omega \left( \frac{\partial J}{\partial \eta } \right)+V\left( \frac{\partial {{P}_{0}}}{\partial \eta } \right).
\end{align}

When the mass $M_0$ of per unit volume of black string is assigned a specific value, Eq. (\ref{3.7}) is reduced to
\begin{align}\label{3.17}
0&={{\left( \frac{\partial {{M}_{0}}}{\partial \eta } \right)}_{{{P}_{0}},S(\eta ),Q(\eta ),J(\eta )}}+{{\left( \frac{\partial M}{\partial S} \right)}_{{{P}_{0}},Q(\eta ),J(\eta )}}{{\left( \frac{\partial S}{\partial \eta } \right)}_{{{P}_{0}},Q(\eta ),J(\eta ),M}}+{{\left( \frac{\partial M}{\partial Q} \right)}_{{{P}_{0}},S(\eta ),J(\eta )}}{{\left( \frac{\partial Q}{\partial \eta } \right)}_{{{P}_{0}},S(\eta ),J(\eta ),M}} \notag \\
&+{{\left( \frac{\partial M}{\partial J} \right)}_{{{P}_{0}},S(\eta ),Q(\eta )}}{{\left( \frac{\partial J}{\partial \eta } \right)}_{{{P}_{0}},S(\eta ),Q(\eta ),M}}+{{\left( \frac{\partial M}{\partial {{P}_{0}}} \right)}_{S(\eta ),Q(\eta ),J(\eta )}}{{\left( \frac{\partial {{P}_{0}}}{\partial \eta } \right)}_{S(\eta ),Q(\eta ),J(\eta ),M}} \notag \\
&={{\left( \frac{\partial {{M}_{0}}}{\partial \eta } \right)}_{{{P}_{0}},S(\eta ),Q(\eta ),J(\eta )}}+T{{\left( \frac{\partial S}{\partial \eta } \right)}_{{{P}_{0}},Q(\eta ),J(\eta ),M}}+U{{\left( \frac{\partial Q}{\partial \eta } \right)}_{{{P}_{0}},S(\eta ),J(\eta ),M}} \notag \\
&+\Omega {{\left( \frac{\partial J}{\partial \eta } \right)}_{{{P}_{0}},S(\eta ),Q(\eta ),M}}+V{{\left( \frac{\partial {{P}_{0}}}{\partial \eta } \right)}_{S(\eta ),Q(\eta ),J(\eta ),M}}.
\end{align}
Eq. (\ref{3.17}) with distinct constraints is provided below. It is observed that when the state parameters of $M$ are represented by $Q, J, S({\eta})$ (note that $Q$ demonstrates the absence of effect of $\eta$ on the charge $Q(\eta)$, the remaining state parameters are analogous), Eq. (\ref{3.17}) is reduced to
\begin{align}\label{3.18}
{{\left( \frac{\partial {{M}_{0}}}{\partial \eta } \right)}_{{{P}_{0}},S(\eta ),Q,J}}&=\underset{M\to {{M}_{0}}}{\mathop{\lim }}\,-T{{\left( \frac{\partial S}{\partial \eta } \right)}_{{{P}_{0}},Q,J,M}},
\end{align}
the thermodynamic relationship of between energy and entropy is expressed in Eq. (\ref{3.18}).

When the state parameters of $M$ are represented by $J({\eta}), S, Q$, Eq. (\ref{3.17}) is reduced to
\begin{align}\label{3.19}
{{\left( \frac{\partial {{M}_{0}}}{\partial \eta } \right)}_{{{P}_{0}},S,Q,J(\eta )}}&=-\underset{M\to {{M}_{0}}}{\mathop{\lim }}\,\Omega {{\left( \frac{\partial J}{\partial \eta } \right)}_{{{P}_{0}},S,Q,M}},
\end{align}
the thermodynamic relationship between energy and angle momentum is expressed in Eq. (\ref{3.19}).

When the state parameters of $M$ are represented by $J, S, Q({\eta})$, Eq. (\ref{3.17}) is reduced to
\begin{align}\label{3.20}
{{\left( \frac{\partial {{M}_{0}}}{\partial \eta } \right)}_{{{P}_{0}},S,Q(\eta ),J}}&=-\underset{M\to {{M}_{0}}}{\mathop{\lim }}\,U{{\left( \frac{\partial Q}{\partial \eta } \right)}_{{{P}_{0}},S,J,M}},
\end{align}
the thermodynamic relationship between energy and charge is expressed in Eq. (\ref{3.20}).

In the event that the state parameters of $M$ are represented by $J, S, Q$ and $M=M_0$ is a specific value, due to the variation of $P_0$ with the perturbation parameter $\eta$, $P_0$ is rewritten as $P_0(\eta)$. Therefore, Eq. (\ref{3.17}) is reduced to
\begin{align}\label{3.21}
{{\left( \frac{\partial {{M}_{0}}}{\partial \eta } \right)}_{{{P}_{0}},S,Q,J}}&=-\underset{M\to {{M}_{0}}}{\mathop{\lim }}\,V{{\left( \frac{\partial {{P}_{0}}(\eta )}{\partial \eta } \right)}_{Q,S,J,M}},
\end{align}
the thermodynamic relationship between energy and pressure is expressed in Eq. (\ref{3.21}).

When the state parameters of $M$ are represented by $Q({\eta}), J({\eta}), S({\eta}), P_{0}{({\eta})}$, Eq. (\ref{3.17}) is reduced to
\begin{align}\label{3.22}
{{\left( \frac{\partial {{M}_{0}}}{\partial \eta } \right)}_{{{P}_{0}},S(\eta ),Q(\eta ),J(\eta )}}&=\underset{M\to {{M}_{0}}}{\mathop{\lim }}\,-T{{\left( \frac{\partial S}{\partial \eta } \right)}_{{{P}_{0}},Q(\eta ),J(\eta ),M}}-\underset{M\to {{M}_{0}}}{\mathop{\lim }}\,\Omega {{\left( \frac{\partial J}{\partial \eta } \right)}_{{{P}_{0}},S(\eta ),Q(\eta ),M}} \notag \\
&-\underset{M\to {{M}_{0}}}{\mathop{\lim }}\,U{{\left( \frac{\partial Q}{\partial \eta } \right)}_{{{P}_{0}},S(\eta ),J(\eta ),M}}-\underset{M\to {{M}_{0}}}{\mathop{\lim }}\,V{{\left( \frac{\partial {{P}_{0}}}{\partial \eta } \right)}_{Q,S,J,M}}
\end{align}
The thermodynamic relationship between energy and the joint action of entropy, angular momentum, charge, and pressure is expressed in Eq. (\ref{3.22}). The Goon and Penco relation for rotating charged black strings, under different constraints, is obtained from Eqs. (\ref{3.18})-(\ref{3.22}).

\section{Discussion}\label{four}

In this study, the thermodynamic relationship associated with the thermodynamic quantities of Rotating Charged Black Strings is investigated in the context of the cosmological constant subjected by the perturbation parameter $eta$. Rotating Charged Black Strings described in cylindrical coordinates are analyzed. The mass of per unit height black strings $M$ at the horizon is served as state parameters determined by angular momentum $J$, charge $Q$, entropy $S$, and pressure $P$, as demonstrated in Eq. (\ref{3.8}). It is imperative to note that when considering spacetime perturbation, denoted by $\eta$ in Eq. (\ref{3.4}), it is evident that under the condition of fixed Black String mass of per unit height, $M=M_0$, the angular momentum, $J$, charge, $Q$, entropy, $S$, and pressure, $P$, exhibit a functional dependence on the perturbation parameter, $eta$.

This finding stands in contrast to the extant theoretical models \cite{M13784,J17014,S115279}. In extant studies, either charge and angular momentum are assumed to be independent of $\eta$, or only a single parameter among the multiple parameters is chosen as a function of $\eta$. It is noteworthy that in rudimentary black hole models, analytical expressions for $\eta$ as a function of each thermodynamic quantity ($M$, $S$, $Q$, etc.) can typically be derived due to the limited number of parameters and the relatively uncomplicated form of the mass function. However, for complex spacetime systems, such as Rotating Charged black strings, the unavailability of explicit analytic expressions for $\eta$ with respect to each thermodynamic quantity poses a significant theoretical challenge in establishing the Goon-Penco universal relation.

The present study proposes a universal expression, Eq. (\ref{3.17}), for the variation of spacetime parameters with $\eta$ based on the first law of thermodynamics for black holes (black strings) in order to address this key scientific issue. Through a systematic examination of the expressions of this equation under various constraints, we have successfully derived the differential relational Eq. (\ref{3.18}) for the variation of state parameters ($S$, $T$, $M$) with $\eta$. This theoretical breakthrough not only provides a novel research paradigm for studying the Goon-Penco universal relation, but more importantly, it opens up new research paths for a deeper understanding of the Weak Gravity Conjecture (WGC). These significant findings not only expand our comprehension of quantum gravity theory, but also furnish substantial theoretical underpinnings and research directions for future studies in this domain.

\section*{Acknowledgments}
We would like to thank Prof. Ren Zhao for their indispensable discussions and comments. This work was supported by the Natural Science Foundation of China (Grant No. 12375050, Grant No. 11705106, Grant No. 12075143), the Scientific Innovation Foundation of the Higher Education Institutions of Shanxi Province (Grant Nos. 2023L269), Shanxi Provincial Natural Science Foundation of China (Grant No. 202203021221211)

\end{document}